\documentclass[twocolumn,showpacs,preprintnumbers,pra,floatfix,amsmath,amssymb]{revtex4}
\usepackage{graphicx}% Include figure files
\usepackage[latin1]{inputenc}
\usepackage[bf,SL,BF]{subfigure}

\newcommand{\ket}[1]{\left| #1 \right>}

\begin{document}

\title{Robust states in semiconductor quantum dot molecules}
\author{H. S. Borges, L. Sanz, J. M. Villas-Bôas and A. M. Alcalde}
\affiliation{Instituto de Física, Universidade Federal de
Uberlândia, 38400-902, Uberlândia-MG, Brazil}

\begin{abstract}
Semiconductor quantum dots coherently driven by pulsed laser are
fundamental physical systems which allow
studying the dynamical properties of confined quantum states.
These systems are attractive candidates for a solid-state qubit, which
open the possibility for several
investigations in quantum information processing. In this work we
study the effects of a specific decoherence process, the spontaneous
emission of excitonic states, in a quantum dot molecule. We model our
system considering a three-level Hamiltonian and solve the
corresponding master equation in the Lindblad form. Our results show
that the spontaneous emission associated with the direct exciton helps
to build up a robust indirect exciton state. This robustness against
decoherence allows potential applications in quantum memories and
quantum gate architectures. We further investigate several regimes of
physical parameters, showing that this process is easily controlled by
tuning of external fields.

\end{abstract}
 \maketitle

\section{Introduction}
\label{sec:intro}

The advance on the manipulation and dynamical control of quantum states under the action of coherent
radiation has recently become a subject of intense research in condensed matter physics.
Dynamical control is a necessary step for the implementation of any protocol associated
with Quantum Information Processing (QIP)~\cite{DiVincenzo00}. In this sense, semiconductor
quantum dots (QDs) has been proved to be an ideal candidate.
Using strong resonant laser pulses and different probe techniques, several different
groups have successfully demonstrated coherent manipulation of the exciton population
of a single QD.~\cite{Stievater:133603,Kamada:246401,Htoon:087401, Zrenner:612,Wang:035306,Stufler:121301}
They demonstrated a process known as Rabi oscillation which is indeed a proof of the
exciton qubit rotation. Unlike atoms, however, QDs suffer from unavoidable variation in
their size, and the presence of a surrounding environment with which they may interact
strongly, making the entire system to lose its phase
quickly~\cite{Vasanelli:216804,Forstner:127401,Villas-Boas:057404}.
The main interest in QDs arises from their characteristic discrete energy spectrum,
and its great flexibility in change it, not only by manipulation of their geometric structure,
but also with the application of external gates. A natural next step for the development
of such system is to put two quantum dots together and allowing them to couple.
A lot of work has been done in this direction, where beautiful examples of a molecule
formation have been
achieved \cite{Bayer:451,Borri:267401,Talalaev:284,Rontani:085327,Krenner:057402, Batteh:155327,Bardot:035314,Ortner:125335,Krenner:184,Unold:137404,Stinaff:636,Robledo2008}. However,
the coherent dynamics of such objects under strong laser pumping remains largely unexplored experimentally,
and our work can give further insight to help the experimental development.

In this paper, we study the effects of the spontaneous decay in the
excitonic states of a self-assembled semiconductor quantum dot
molecule (QDM) coupled by tunneling and under the influence of an
external electromagnetic field. We use a standard density matrix
approach in the Lindblad form to describe the system dynamics and
our results indicate that the spontaneous decay of the direct
exciton helps to build up a coherent population of the indirect
exciton (electron and hole in different dots), which has a longer
lifetime due to its spatial separation with small overlap of the
wave function. This effect is robust is robust to the changes of external parameters and
in order to describe it we describe the physical system and the detailed
theoretical model in Sec.~\ref{sec:theory}, then in
Sec.~\ref{sec:results} we show the results of numerical
calculations, followed by our conclusions in Sec.~\ref{sec:summary}.

\section{Theory and model}
\label{sec:theory}
%The system is studying through the three-level
%model, with $\ket{0}$ state corresponding to the vacuum; the
%$\ket{1}$ being the direct exciton state (which is defined as
%electron and hole in the same dot) and $\ket{2}$ represents an
%indirect exciton (the electron and hole are in different quantum
%dots).

\begin{figure}[htbp]
\begin{center}
\includegraphics[scale=0.7]{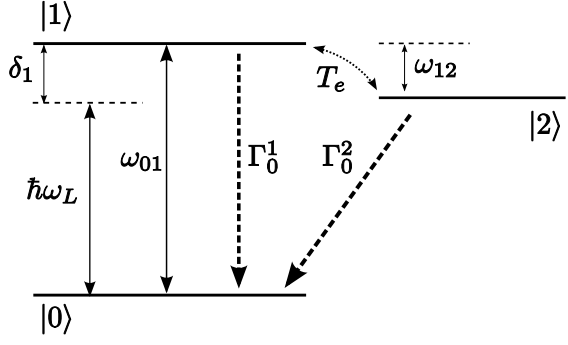}
\end{center}
\caption{Scheme of energy levels with physical parameters on Hamiltonian~(\ref{eq:H}).}
\label{fig:levels}
\end{figure}

The physical system we consider here is an asymmetric double quantum
dot coupled by tunneling. Electrons and holes can be confined in
either dot and we can use a near-resonant optical pulse to promote
electrons from the valence to the conduction band, creating an
electron-hole correlated state known as exciton. Electrons or holes
can then tunnel to the other dot, creating an indirect exciton.
An external electric field, applied in the growth direction, brings the individual levels
of electrons or holes into resonance, favoring the tunneling.
Nevertheless, in asymmetric QDM structures it is even possible to control which type
of carrier, electron or holes, tunnels~\cite{Bracker06}.
In this situation, we can safely neglect the tunneling of holes as the
electric field brings one level (conduction band)
more into resonance while makes the other (valence band) more out of
resonance.
With this assumption, the dynamics of the QDM can be modeled by a
simple three-level system, where the ground state $|0\rangle$ is a
molecule without any excitation, $|1\rangle$ is the system with one
exciton in the left dot, while $|2\rangle$ is the system with one
indirect exciton, after the electron has tunneled.
%The states $|1\rangle$ and $|2\rangle$ and then coupled by tunneling of one electron $T_e$.
The schematic configuration of levels and
physical parameters are shown in the Fig.~\ref{fig:levels}, where we also include the decoherence channels
associated with spontaneous emission of excitonic
states ($\Gamma^{1}_{0}$, $\Gamma^{2}_{0}$).

Using the rotating wave approximation and dipole approximation, the system Hamiltonian is written as ~\cite{Villas04}:
\begin{eqnarray}
\label{eq:H} \hat{H}(t)&
=\sum_{j=0}^{2}\hbar\omega_{j}|j\rangle\langle
j|+T_{e}(|1\rangle\langle2|+ |2\rangle\langle 1|)\nonumber
\\&+\hbar\Omega (e^{i\omega_L t}|0\rangle\langle1|+e^{-i\omega_L
t}|1\rangle\langle0|),
\end{eqnarray}
where $\omega_{j}$ are the frequencies of $|j\rangle$-th states ($j=0,1,2$), $T_{e}$ is
the tunneling coupling, $\omega_L$ is the frequency of the applied laser and the dipole coupling is
%the strength of
%light-matter interaction, which depends on the applied electric
%field and electronic structure of the semiconductor. In this system,
%the wavelength of the coupled photon mode is typically larger than
%the spatial extension of the corresponding electron density of
%probability. In front of this, the dipole approximation is valid and
%$\Omega$ is written as
$\Omega = \langle0|\overrightarrow{\mu}
\cdot\overrightarrow{E}|1\rangle/2\hbar$, where
$\overrightarrow{\mu}$ is the electric dipole moment and
$\overrightarrow{E}$ is the amplitude of incident field.
%The Rabi frequency is
%proportional to the amplitude of the incident electromagnetic pulse
%$\overrightarrow{E}(t)$.
The intensity of incident field can be easily controlled to provide available conditions
for the coherent control of the system quantum state.
We also assume a low-intensity incident pulse, so the Rabi frequency is significatively
smaller than the intraband excitation energy ~\cite{Calarco03}, $\Omega \ll \omega_{10}=\omega_1-\omega_0$
and $\omega \approx \omega_{10}$.
Under this assumption, we might consider that only the ground-state exciton can be formed in our system.

Applying the unitary transformation~\cite{Villas04}
\begin{equation}
\label{eq:U} \hat{U}=\exp \left[\frac{i\omega_L
t}{2}\left(|1\rangle\langle1|-|0\rangle\langle0|+|2\rangle\langle2|
\right)\right],
\end{equation}
and using the \emph{Baker-Hausdorff lemma}~\cite{Sakurai94} , we obtain a time-independent version of Hamiltonian (\ref{eq:H}) written as follows:
\begin{equation}
\label{eq:Htras}
\hat{{H}}^{'}=\frac{1}{2}\left(\begin{array}{ccc}-\delta_{1}&2\hbar\Omega&0\\
2\hbar\Omega&\delta_{1}&2T_e\\0&2T_e&\delta_{2}
\end{array}\right),
\end{equation}
where $\delta_{1}=\hbar\left(\omega_{10}-\omega_L\right)$ is the detuning between the frequency
of optical pulse and exciton transition, $\delta_{2}=\delta_{1}+2\hbar\omega_{21}$ and $\omega_{ij}$ is
the optical transition between $i$ and $j$ energy states.

%The dynamics of this model exhibits more complexity if we
%compare with the two-level system: it can be solved analytically
%only when the frequencies are resonant
%($\omega_{10}=\omega_L=\omega_{21}$) so $\delta_1=0$ and $\delta_2=0$.
%For this choice, it is straightforward to find the eigenstates and
%eigenvalues of Hamiltonian (\ref{eq:H})~\cite{Villas04}. Other
%choices of physical parameters require numerical
%procedures in order to solve time-independent Schrödinger equation.

To taking into account the effects of decoherence, we used the Liouville-Von Neumman-Lindblad equation given by~\cite{Zhao03}:
\begin{equation}\label{eq:master}
\frac{\partial{\hat{\rho}\left(t\right)}}{\partial t}=-\frac{i}{\hbar}[\hat{H},\hat{\rho}\left(t\right)]+\hat{L}(\hat{\rho}\left(t\right)).
\end{equation}
Here, $\hat{\rho}\left(t\right)$
%$\hat{\rho}\left(t\right)=\sum_{k}p_{k}|\psi_{k}\left(t\right)\rangle\langle\psi_{k}(t)|$
is the density matrix operator.
%with $p_{k}$ being the probability of
%the system to be in  $|\psi_{k}(t)\rangle$ state at time $t$.
The Liouville operator, $\hat{L}(\hat{\rho})$, describes the dissipative process. Assuming the Markovian approximation, Liouville operator can be written as~\cite{Villas07}:
\begin{equation}\label{eq:liouville}
\hat{L}(\hat{\rho})= \frac{1}{2} \sum_{i}\Gamma^i_j (2|j\rangle\langle i|\hat{\rho}|i\rangle\langle j|-\hat{\rho}|i\rangle\langle i|-|i\rangle\langle i|\hat{\rho}),
\end{equation}
where $\Gamma^i_j$ corresponds to the decoherence rates due spontaneous decay from the state $|i\rangle$ to the state $|j\rangle$. In order to investigate the dynamics associated with this physical system, we solve the master equation ~(\ref{eq:master}), and found the density matrix coefficients at certain time $t$. Writing Eq.(\ref{eq:master}) in the basis defined by $|0\rangle$, $|1\rangle$ and $|2\rangle$ states, we obtain a set of nine coupled linear differential equations written as:
\begin{eqnarray}
\label{eq:rhodot}
\dot{\rho}_{00}&=&-i\Omega(\rho_{10}-\rho_{01})+\Gamma^{1}_{0}\rho_{11} +\Gamma^{2}_{0}\rho_{22},\nonumber\\
\dot{\rho}_{01}&=&\frac{i}{\hbar}\left[\delta_{1}\rho_{01}+\hbar\Omega(\rho_{00}-\rho_{11})+T_e\rho_{02}\right]-\frac{1}{2}\Gamma^{1}_{0}\rho_{01},\nonumber\\
\dot{\rho}_{02}&=&\frac{i}{\hbar}\left[\frac{\rho_{02}}{2}(\delta_{1}+\delta_{2})-\hbar\Omega\rho_{12}+T_e\rho_{01}\right]-\frac{1}{2}\Gamma^{2}_{0}\rho_{02},\nonumber\\
\dot{\rho}_{10}&=&\frac{i}{\hbar}\left[-\delta_{1}\rho_{10}+\hbar\Omega(\rho_{11}-\rho_{00})-T_e\rho_{20}\right]-\frac{1}{2}\Gamma^{1}_{0}\rho_{10},\nonumber\\
\dot{\rho}_{11}&=&\frac{i}{\hbar}\left[\hbar\Omega(\rho_{10}-\rho_{01})+T_e(\rho_{12}-\rho_{21})\right]-\Gamma^{1}_{0}\rho_{11},\nonumber\\
\dot{\rho}_{12}&=&\frac{i}{\hbar}\left[\frac{\rho_{12}}{2}(\delta_{2}-\delta_{1})-\hbar\Omega\rho_{02}+T_e(\rho_{11}-\rho_{22})\right]\nonumber\\&&-\frac{1}{2}(\Gamma^{1}_{0}+\Gamma^{2}_{0})\rho_{12},\nonumber\\
\dot{\rho}_{20}&=&\frac{i}{\hbar}\left[-\frac{\rho_{20}}{2}(\delta_{2}+\delta_{1})+\hbar\Omega\rho_{21}-T_e\rho_{10}\right]-\frac{1}{2}\Gamma^{2}_{0}\rho_{20},\nonumber\\
\dot{\rho}_{21}&=&\frac{i}{\hbar}\left[\frac{\rho_{21}}{2}(\delta_{1}-\delta_{2})+\hbar\Omega\rho_{20}+T_e(\rho_{22}-\rho_{11})\right]\nonumber\\&&-\frac{1}{2}(\Gamma^{1}_{0}+\Gamma^{2}_{0})\rho_{21},\nonumber\\
\dot{\rho}_{22}&=&\frac{i}{\hbar}T_e \left(\rho_{21}-\rho_{12}\right)- \Gamma^{2}_{0}\rho_{22}.
\end{eqnarray}

In order to solve the set of equations~(\ref{eq:rhodot}), we rewrite as $\dot{\rho}=A\rho$, considering $\rho$ as a column vector and $A$ being a square matrix associated with the coefficients of the coupled system above.
%That means we must perform a full diagonalization procedure on matrix $A$ to obtain an uncoupled equations system. Once we solve the eigenproblem for $A$, we are able to write the density matrix elements as:
The solution can be written as
\begin{equation}\label{eq:solve}
\rho_{ij}(t)=\sum^{8}_{j=0}S_{ij}e^{\lambda_{j}t}\left( S^{-1}_{ij}\rho_{ij}(0)\right),
\end{equation}
where, $\lambda_{j}$ and $S_{ij}$ are the eigenvalues and the matrix formed by the eigenvectors of matrix $A$, respectively. $\rho_{ij}(0)$ are the elements of the density matrix operator at $t=0$.

\section{Results and Discussion}\label{sec:results}
%Here, we will focuss our discussion on the dynamics associated with a quantum dot molecule.
%We are interested
%on the effect of spontaneous emission of direct and indirect
%excitons. This decoherence process is studied in details performing
%numerical calculations of the average population of indirect exciton
%state ($\ket{2}$ state) as function of the coupling parameters
%$\Omega$ and $T_e$ and decoherence parameters $\Gamma^i_j$ on
%Eq.(\ref{eq:master}).
For our calculations, we consider the following values of physical
parameters: $\hbar\omega_{10}\simeq 1.6$ eV~\cite{Kamada:246401,Gammon98},
$\Omega\simeq 0.05-1.0$ meV~\cite{Chen01,Calarco03}, $\Gamma^{1}_{0}\simeq$ 0.33 - 6.6 $\mu$eV~\cite{Chen01,Takagahara02}, and $\Gamma^{2}_{0}\simeq 10^{-4}\Gamma^{1}_{0}$~\cite{Negoita99}.
The tunneling coupling, which depends the barrier characteristics and the external
electric field, was selected as: $T_e\simeq 0.01-0.1$ meV~\cite{Tackeuchi00} or $T_e\simeq 1-10$ meV~\cite{Emary07},
for weak and strong tunneling regime, respectively. The system dynamics depends also
from the detunings $\delta_1$ and $\delta_2$. Experimentally, $\delta_1$ is controlled
by varying the frequency of external laser. The value of $\delta_2$ is changed by varying
$\delta_1$ and the frequency transition $\omega_{21}$, which can be done by manipulation of external electric
field that changes the effective confinement potential. By varying this set of parameters we are
able to perform a coherent manipulation of the wave function of the system.
%We scale all our
%parameters in units of $\omega=1$ meV.
For all simulations, we consider $|\Psi(0)\rangle=|0\rangle$ as initial condition.

\begin{figure}[htbp]
%\begin{center}
\includegraphics[scale=0.55]{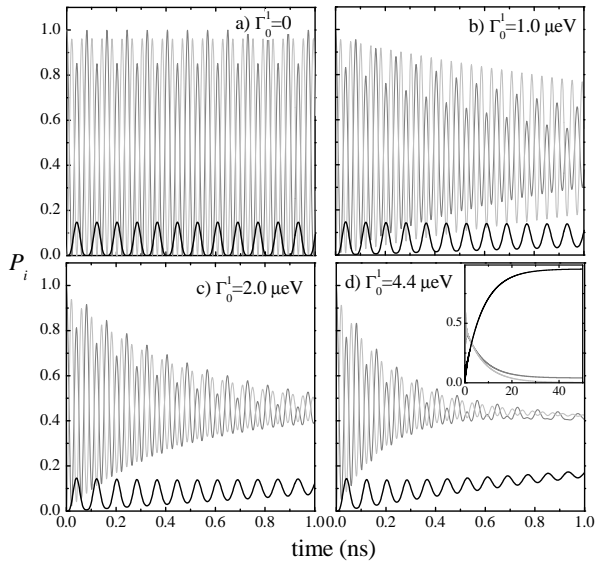}
%\end{center}
\caption{Dynamics of populations, $P_i$, of level $\ket{i}$ ($i=0,1,2$) as function of time for
different choices of spontaneous emission rate $\Gamma^1_0$ with parameters $\delta_1=0$, $\omega_{21}=0$,
$\Omega$=50 $\mu$eV and $T_e$=10 $\mu$eV.
We use gray line for $P_0$, light-gray line for $P_1$ and black line for $P_2$.
(a) Non-dissipative dynamics ($\Gamma^1_0=0$);
(b)$\Gamma^1_0$=1.0 $\mu$eV;
(c)$\Gamma^1_0$=2.0 $\mu$eV;
(d)$\Gamma^1_0$=4.4 $\mu$eV.} \label{fig:pop}
\end{figure}
%We will show that we can distinguish two different behaviors of
%population dynamics depending on the physical parameters of our
%problem: one is the Rabi oscillation regime and the second is
%characterized by stationary populations, which appears when the
%effect of spontaneous emission is considered.
Our first task is to analyze the effect on population dynamics of the decoherence process
associated with spontaneous emission of direct exciton. In Fig.~\ref{fig:pop},
we plot the probability of occupation associated with each of the three levels considering
different values of spontaneous emission rate $\Gamma^1_0$.
For the physical parameters considered here, the non-dissipative dynamics ($\Gamma^1_0=0$) shows
that there are Rabi oscillations between the three levels of the system.
This is illustrated in Fig.~\ref{fig:pop} a).
The population of indirect exciton, state $\ket{2}$, depends directly on the parameter $T_e$,
although the value of the coupling $\Omega$ and detunings $\delta_1$ and $\omega_{21}$, has important
effects on dynamics~\cite{Villas04}. The situation changes when spontaneous emission is taken into account.
As we expected, the Rabi oscillations become damped. This can be seen in Figs.~\ref{fig:pop} b), c) and d).
For long times and values of $\Gamma^1_0$ high enough, as shown in the inset of Fig.~\ref{fig:pop} d),
the Rabi oscillations are suppressed
% even at short times.
%At long times,
and the electron wave function tends to an asymptotic state.
%This asymptotic state depends on the physical parameters of Hamiltonian.
%For example, we can
%obtain a wave function so $P_0=P_2=0.5$, indicating a state that
%keeps similarities with a dark state defined for atomic
%states~\cite{RevEIT}.
%We will discuss the dependence of asymptotic
%states and physical parameters of Hamiltonian (\ref{eq:H}) in a
%future work.

%Second issue is to understand the behavior of average occupation of level $\ket{2}$, varying the coupling parameters $\Omega$ and $T_e$ and considering different values of spontaneous emission $\Gamma^{1}_{0}$.
Now we focuss our attention on the formation of a stationary state with high population of
indirect excitonic level, $\ket{2}$.
With a lifetime significatively longer (about $10^4$ times the direct exciton)~\cite{Negoita99},
this particular state shows more potential for quantum information processing than the
direct exciton, $\ket{1}$ state.
In order to study the effects of several physical parameters on Hamiltonian (\ref{eq:H}) and
the decoherence, we study the behavior of average occupation of state $\ket{2}$, defined as
\[\overline{P_2}=\frac{1}{t_{\infty}}\int^{t_{\infty}}_{0}P_2\left(t\right)dt.\]
%with $t_{\infty}\gg t_R$ being $t_R=1/\sqrt{\Omega^2+T_e^2}$ the period of the Rabi oscillations.
%High values of the quantity $\overline{P_2}$ indicate high values of occupation of level $\ket{2}$.
\begin{figure}[htbp]
%\begin{center}
\includegraphics[scale=0.46]{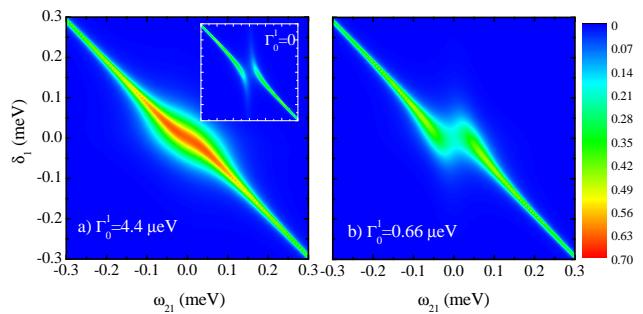}
%\end{center}
\caption{(Color online) Average population of state $\ket{2}$,
$\overline{P_2}$, as function of detuning, $\delta_1$ and frequency
$\omega_{21}$ for $\Omega$=50 $\mu$eV, $T_e$=10 $\mu$eV and
$\Gamma^{2}_{0}\approx 10^{-4}\Gamma^{1}_{0}$.
a)$\Gamma^{1}_{0}$=4.4  $\mu$eV. Inset: $\overline{P_2}$ for
$\Gamma^{1}_{0}=0$ and the same values of $\Omega$ and $T_e$.
b)$\Gamma^{1}_{0}$=0.66 $\mu$eV. } \label{fig:P2det}
\end{figure}

In Fig.~\ref{fig:P2det}, we plot our results for $\overline{P_2}$, as function of laser
detuning $\delta_1$ and frequency $\omega_{21}$, considering two different values of the direct exciton
spontaneous emission rate, $\Gamma^{1}_{0}$.
Bright colors are associated with high values of $\overline{P_2}$, which means an
efficient transference of the electron from the first to the
second dot.
From our results, it is possible to conclude that a large occupation probability
of $\ket{2}$  is obtained if the detuning $\delta_1$ is
balanced with the applied electric field so that $\delta_1+\omega_{21}\simeq 0$.
We will named this condition as \emph{balanced detuning}.
The behavior considering full resonance between the three levels ($\delta_1\simeq\omega_{21}\simeq 0$) deserves
more attention.
Let us define an area associated with the full resonance
condition $\left|\delta_1\right|,\left|\omega_{21}\right|\lesssim 50 \mu$eV:
%marked by a square with dashed lines in Fig.~\ref{fig:P2det}.
when spontaneous emission is not considered ($\Gamma^1_0$=0), the average population $\overline{P_2}$ is near to zero, as shown in Ref.~\cite{Villas04} and in the
inset of Fig.~\ref{fig:P2det} a).
Thus, full resonance condition is not a good experimental choice
for an optimal creation of indirect excitonic state.
Considering the effects of spontaneous emission $\Gamma^{1}_{0}$, we can observe a different behavior:
the values of $\overline{P_2}$ at point $\left(\omega_{21},\delta_1\right)=\left(0,0\right)$ increase
from $0.05$ (for $\Gamma^{1}_{0}=0$) to $\simeq 0.6$ (for $\Gamma^{1}_{0}$=4.4 $\mu$eV) and $\simeq 0.2$ (for $\Gamma^{1}_{0}$=0.66 $\mu$eV).
This shows that for realistic direct excitons, with a non-zero spontaneous emission rate,
the transfer of the electron between dots is more efficient.
\begin{figure}[htbp]
\begin{center}
\includegraphics[scale=0.55]{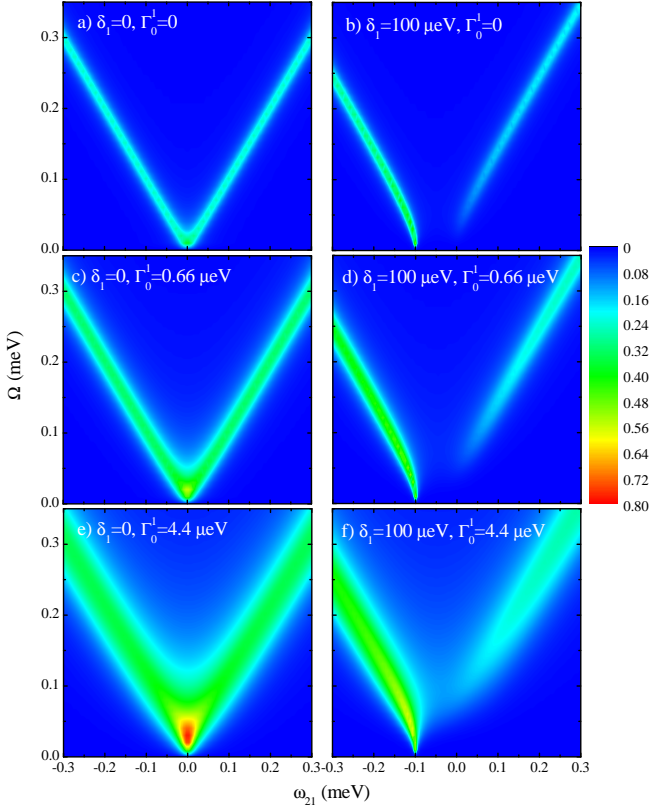}
\end{center}
\caption{(Color online). Average population of state $\ket{2}$,
$\overline{P_2}$, as function of coupling parameter $\Omega$, and
frequency $\omega_{21}$ for resonant and non-resonant condition.
a) $\Gamma^{1}_{0}=0$ and $\delta_1=0$;
b) $\Gamma^{1}_{0}=0$ and $\delta_1$=100 $\mu$eV;
c) $\Gamma^{1}_{0}$=0.66 $\mu$eV and $\delta_1=0$;
d) $\Gamma^{1}_{0}$=0.66 $\mu$eV and $\delta_1$=100 $\mu$eV;
e) $\Gamma^{1}_{0}$=4.4 $\mu$eV and $\delta_1=0$;
f) $\Gamma^{1}_{0}$=4.4 $\mu$eV and $\delta_1$=100 $\mu$eV.
In all cases, $T_e$=10 $\mu$eV and
$\Gamma^{2}_{0}\approx 10^{-4}\Gamma^{1}_{0}$ }
\label{fig:P2G}
\end{figure}

In Fig.~\ref{fig:P2G}, we show our results for average population, $\overline{P_2}$, as a
function of both, frequency $\omega_{21}$ and dipole coupling $\Omega$, for different
choices of $\Gamma^{1}_{0}$ considering $\delta_1=0$, Figs.~\ref{fig:P2G} a), c) and e),
and $\delta_1$=100 $\mu$eV, Figs.~\ref{fig:P2G} b), d) and f). For all cases, we are able
to populate the indirect exciton state, evidenced by bright regions with values of
$\overline{P_2}$ larger than $0.3$. At resonance condition, Figs.~\ref{fig:P2G} a), c) and
e), this bright area have a \textsf{V}-like form, with higher values of $\overline{P_2}$
concentrated on a small area associated with low values of $\Omega$ and $\omega_{21}$.
For non-resonant condition, the symmetry between negative and positive values of
$\omega_{21}$ is broken. Still, the large values of $\overline{P_2}$ are obtained when the
condition $\delta_1+\omega_{21}\simeq 0$ is fulfilled. The action of spontaneous emission
can be analyzed by comparing the different situations shown in Fig.~\ref{fig:P2G}. Higher
values of parameter $\Gamma^{1}_{0}$ are connected with higher values of average
population $\overline{P_2}$. For example, in Fig.~\ref{fig:P2G} a) when
$\Gamma^{1}_{0}=0$ the maximum value of $\overline{P_2}\simeq 0.36$. Considering
decoherence, the maximum value of $\overline{P_2}$ goes from $0.6$ for
$\Gamma^{1}_{0}$=0.66 $\mu$eV [Fig.~\ref{fig:P2G} c)] to $\simeq 0.8$ for
$\Gamma^{1}_{0}$=4.4 $\mu$eV [Fig.\ref{fig:P2G} d)]. Also, the total area for highly
efficient population of $\ket{2}$ state increase as spontaneous emission increase: both,
the arms of the characteristic \textsf{V} area and the region with best values of
$\overline{P_2}$ become progressively large when the value of $\Gamma^{1}_{0}$ increase.

\begin{figure}[htbp]
%\begin{center}
\includegraphics[scale=0.5]{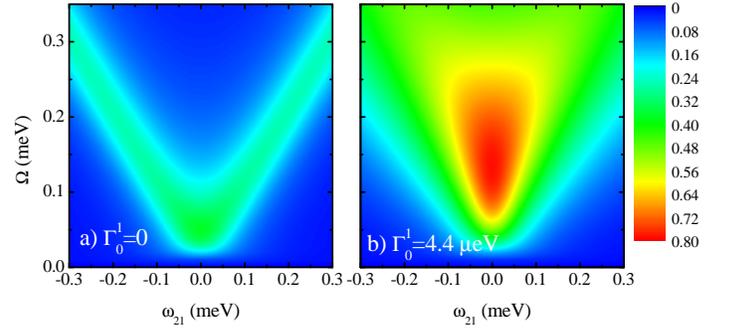}
%\end{center}
\caption{(Color online). Average occupation of state $\ket{2}$, $\overline{P_2}$, as
function of coupling parameter, $\Omega$, and frequency $\omega_{21}$
considering $T_e$=50 $\mu$eV, $\delta_1=0$ and $\Gamma^{2}_{0}\approx 10^{-4}\Gamma^{1}_{0}$.
(a) $\Gamma^{1}_{0}=0$.
(b) $\Gamma^{1}_{0}$=4.4 $\mu$eV.}
\label{fig:P2Te}
\end{figure}

It is useful to check the combined effect of both, the tunneling and decoherence.
It is expected a good transfer of population associated with higher values of $T_e$ parameter.
This can be verified by comparing the results $\overline{P_2}$ without the effect of decoherence
process with $T_e$=10 $\mu$eV, Fig.~\ref{fig:P2G} a), with the results considering a higher value
of tunneling parameter $T_e=50\mu$eV, Fig.~\ref{fig:P2Te} a). The effect of decoherence process
is illustrated by Fig.~\ref{fig:P2Te} b). Notice that area on Fig.\ref{fig:P2Te} with
high $\overline{P_2}$ increase by the action of decoherence and the maximum
value of $\overline{P_2}$ goes from $\simeq 0.4$, in Fig.~\ref{fig:P2Te} a), to $\overline{P_2}\simeq 0.8$ in
Fig.~\ref{fig:P2Te} b).

\begin{figure}[tbp]
%Arquivo origin: ./figs paper/compop.OPJ feito a partir dos dados de rabitosta e os arquivos localizados em./figs paper/compop
%\begin{center}
\includegraphics[scale=0.53]{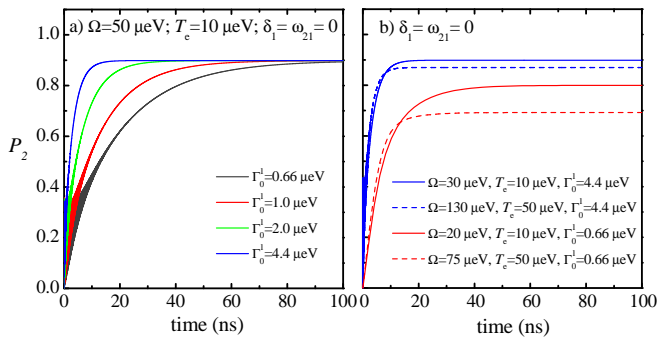}
%\end{center}
\caption{(Color online) Population of indirect exciton state, $P_2$, as function of time at
stationary regime.
(a)$P_2$ for different values of spontaneous emission
rate $\Gamma^1_0$ considering full resonance
condition ($\delta_1=0$, $\omega_{21}=0$) for coupling parameters $\Omega$=50 $\mu$eV and $T_e$=10 $\mu$eV.
(b) $P_2$ associated with the physical parameters for the highest
values of $\overline{P_2}$ founded in Fig.\ref{fig:P2G} and Fig.\ref{fig:P2Te}.} \label{fig:compop}
\end{figure}

After our analysis of $\overline{P_2}$, it is necessary to check the actual behavior of level
population $P_2$. In Fig.\ref{fig:compop}, we plot $P_2$ considering some choices of physical
parameters associated with our previous analysis (Figs.~\ref{fig:pop}~-~\ref{fig:P2Te}).
In all cases, we limit ourselves to full resonance condition ($\delta_1\simeq\omega_{21}\simeq 0$).
When dynamics is associated with stationary states, the value of $\overline{P_2}$ depends on
two aspects: the final value of $P_2$ at stationary state and the time needed to reach this
maximum value.
In Fig.~\ref{fig:compop} a) we plot $P_2$ for $\Omega$=50 $\mu$eV and $T_e$=10 $\mu$eV considering different
values of $\Gamma^1_0$. We can conclude that a higher spontaneous emission rate of the direct exciton is
connected with a faster evolution to the asymptotic value of $P_2$. That means, the broadening
effects (short lifetime) on the direct exciton are advantageous if we are interested on
manipulate electronic wave function in order to
create an asymptotic state with high values of $P_2$ at short times.

Next, we verify that the exact maximum value of $P_2$ is related with coupling
parameters $\Omega$ and $T_e$ and, also, with the balanced detuning
condition $\delta_1+\omega_{21}\simeq0$.
From our calculations of average occupation of
indirect excitonic state, we analise the behavior of $P_2$ considering a set of $\Omega$ and $T_e$ parameters associated with maximum values of $\overline{P_2}$ for two different $\Gamma^1_0$ rates.

The evolution of population $P_2$ is shown in Fig.~\ref{fig:compop} b), blue (red) lines
represent the $\Gamma^1_0$=4.4 $\mu$eV ($\Gamma^1_0$=0.66 $\mu$eV) situation.
%We note that, for all choices, the times at the asymptotic stationary states
%are low, although we note that the lowest time is obtained if we increase one
%of the coupling parameters $T_e$ and $\Omega$.
%This can be check by comparing both, red and blue lines.
%Finally, for a lower spontaneous emission rate, we obtain a lower value of $P_2$ even
%in the condition for the highest value of $\overline{P_2}$.
%This is a signature than this particular asymptotic state is robust to the decoherence process.
%%google
For a fixed value of $\Gamma^1_0$, we can define the
characteristic time $t_0$ as the time at which the system reaches
the asymptotic value of $P_2$ (for example $t_0\simeq$ 14ns for
$\Gamma^1_0$ = 0.66$\mu$eV). This characteristic time, which depends on
the value of $\Gamma^1_0$, allows us to distinguish two dynamical
regimes: 1) for long time, $t\gg t_0$ the population is essentially
independent of time and its maximum value depends directly on the
$\Omega/T_e$ rate value (the population increases when this rate
increases). 2) at short times, $t <t_0$, the dynamics does not
depend on the value of $\Omega/T_e$, being governed mainly by
$\Gamma^1_0$. Comparing all cases plotted in Fig.~\ref{fig:compop} b) we can
conclude that the condition to obtain an asymptotic state with
large values of the occupation $P_2$, associated with short
characteristic times $t_0$, is given by $ \frac{\Omega}{Te} \simeq \frac{Te}{\Gamma^1_0}$.
Thus, it is possible to obtain experimentally optimized values of $P_2$, by adjusting
appropriately the laser intensity $\Omega$ for fixed values of $T_e$ and $\Gamma^1_0$, which in
turn can be obtained through optical spectroscopy.
%% end google

\section{Summary}\label{sec:summary}

In this work, we use a standard density matrix approach in the Lindblad form to model the dynamics of a
Quantum Dot Molecule under the influences of externals electric and electromagnetic fields, and in the
presence of spontaneous emission.
By numerically solving the density matrix we show that the spontaneous decay of the direct
exciton helps to build up a coherent population of the indirect exciton, which should have important
applications in quantum information processing due to its longer coherence time.

We further investigate the efficiency of creation of indirect exciton state as function of physical
parameters of our model.
For weak spontaneous emission rate, the system presents a Rabi
oscillation and in the opposite limit the system rapidly build up a stationary
population of the indirect exciton.
Our results shown that the population of the indirect exciton is strongly influenced
by the spontaneous emission of the direct exciton.
We demonstrate that the indirect exciton, which has a longer lifetime, is robust
against the spontaneous emission process.
Finally, at maximum average population conditions, we determined a relation between the
relevant parameters of the system which allows us to obtain large populations of indirect
exciton $P_2 \simeq$ 0.9.
%%sanz
%We investigate the efficiency of creation of indirect excitonic state by calculating both,
%the population of indirect excitonic level, $P_2$, and the average
%occupation, $\overline{P_2}$, as function of physical parameters of our model.
%First, we compare the dynamics of both cases, with and without decoherence.
%We demonstrate that the system has two distinguishable behaviors, which depends directly
%on spontaneous emission rate.
%First one is a Rabi oscillation behavior, which is governed by couplings $\Omega$ and $T_e$.
%Second one is a stationary behavior, connected with decoherence process, when the Rabi oscillation
%are suppressed at certain critical time.
%Our results shown that the population of the indirect exciton is strongly
%influenced by the spontaneous emission of the direct exciton.
%We demonstrate that the indirect exciton, which has a longer lifetime, is
%robust against the spontaneous emission process.
%%end sanz

\acknowledgments{The authors gratefully acknowledge financial support from Brazilian Agencies CAPES, CNPq and FAPEMIG. This work was performed as part of the Brazilian National
Institute of Science and Technology for Quantum Information
(INCT-IQ).}
%\bibliographystyle{unsrt}
%\bibliography{borges09}

\begin{thebibliography}{35}
\expandafter\ifx\csname natexlab\endcsname\relax\def\natexlab#1{#1}\fi
\expandafter\ifx\csname bibnamefont\endcsname\relax
  \def\bibnamefont#1{#1}\fi
\expandafter\ifx\csname bibfnamefont\endcsname\relax
  \def\bibfnamefont#1{#1}\fi
\expandafter\ifx\csname citenamefont\endcsname\relax
  \def\citenamefont#1{#1}\fi
\expandafter\ifx\csname url\endcsname\relax
  \def\url#1{\texttt{#1}}\fi
\expandafter\ifx\csname urlprefix\endcsname\relax\def\urlprefix{URL }\fi
\providecommand{\bibinfo}[2]{#2}
\providecommand{\eprint}[2][]{\url{#2}}

\bibitem[{\citenamefont{DiVincenzo}(2000)}]{DiVincenzo00}
\bibinfo{author}{\bibfnamefont{D.~P.} \bibnamefont{DiVincenzo}},
  \bibinfo{journal}{Fortschritte der Physik} \textbf{\bibinfo{volume}{48}},
  \bibinfo{pages}{771} (\bibinfo{year}{2000}).

\bibitem[{\citenamefont{Stievater et~al.}(2001)\citenamefont{Stievater, Li,
  Steel, Gammon, Katzer, Park, Piermarocchi, and Sham}}]{Stievater:133603}
\bibinfo{author}{\bibfnamefont{T.~H.} \bibnamefont{Stievater}},
  \bibinfo{author}{\bibfnamefont{X.}~\bibnamefont{Li}},
  \bibinfo{author}{\bibfnamefont{D.~G.} \bibnamefont{Steel}},
  \bibinfo{author}{\bibfnamefont{D.}~\bibnamefont{Gammon}},
  \bibinfo{author}{\bibfnamefont{D.~S.} \bibnamefont{Katzer}},
  \bibinfo{author}{\bibfnamefont{D.}~\bibnamefont{Park}},
  \bibinfo{author}{\bibfnamefont{C.}~\bibnamefont{Piermarocchi}},
  \bibnamefont{and} \bibinfo{author}{\bibfnamefont{L.~J.} \bibnamefont{Sham}},
  \bibinfo{journal}{Phys. Rev. Lett.} \textbf{\bibinfo{volume}{87}},
  \bibinfo{eid}{133603} (\bibinfo{year}{2001}).

\bibitem[{\citenamefont{Kamada et~al.}(2001{\natexlab{a}})\citenamefont{Kamada,
  Gotoh, Temmyo, Takagahara, and Ando}}]{Kamada:246401}
\bibinfo{author}{\bibfnamefont{H.}~\bibnamefont{Kamada}},
  \bibinfo{author}{\bibfnamefont{H.}~\bibnamefont{Gotoh}},
  \bibinfo{author}{\bibfnamefont{J.}~\bibnamefont{Temmyo}},
  \bibinfo{author}{\bibfnamefont{T.}~\bibnamefont{Takagahara}},
  \bibnamefont{and} \bibinfo{author}{\bibfnamefont{H.}~\bibnamefont{Ando}},
  \bibinfo{journal}{Phys. Rev. Lett.} \textbf{\bibinfo{volume}{87}},
  \bibinfo{eid}{246401} (\bibinfo{year}{2001}{\natexlab{a}}).

\bibitem[{\citenamefont{Htoon et~al.}(2002)\citenamefont{Htoon, Takagahara,
  Kulik, Baklenov, A.~L.~Holmes, and Shih}}]{Htoon:087401}
\bibinfo{author}{\bibfnamefont{H.}~\bibnamefont{Htoon}},
  \bibinfo{author}{\bibfnamefont{T.}~\bibnamefont{Takagahara}},
  \bibinfo{author}{\bibfnamefont{D.}~\bibnamefont{Kulik}},
  \bibinfo{author}{\bibfnamefont{O.}~\bibnamefont{Baklenov}},
  \bibinfo{author}{\bibfnamefont{J.}~\bibnamefont{A.~L.~Holmes}},
  \bibnamefont{and} \bibinfo{author}{\bibfnamefont{C.~K.} \bibnamefont{Shih}},
  \bibinfo{journal}{Phys. Rev. Lett.} \textbf{\bibinfo{volume}{88}},
  \bibinfo{eid}{087401} (\bibinfo{year}{2002}).

\bibitem[{\citenamefont{Zrenner et~al.}(2002)\citenamefont{Zrenner, Beham,
  Stufler, Findeis, Bichler, and Abstreiter}}]{Zrenner:612}
\bibinfo{author}{\bibfnamefont{A.}~\bibnamefont{Zrenner}},
  \bibinfo{author}{\bibfnamefont{E.}~\bibnamefont{Beham}},
  \bibinfo{author}{\bibfnamefont{S.}~\bibnamefont{Stufler}},
  \bibinfo{author}{\bibfnamefont{F.}~\bibnamefont{Findeis}},
  \bibinfo{author}{\bibfnamefont{M.}~\bibnamefont{Bichler}}, \bibnamefont{and}
  \bibinfo{author}{\bibfnamefont{G.}~\bibnamefont{Abstreiter}},
  \bibinfo{journal}{Nature} \textbf{\bibinfo{volume}{418}},
  \bibinfo{pages}{612} (\bibinfo{year}{2002}).

\bibitem[{\citenamefont{Wang et~al.}(2005)\citenamefont{Wang, Muller, Bianucci,
  Rossi, Xue, Takagahara, Piermarocchi, MacDonald, and Shih}}]{Wang:035306}
\bibinfo{author}{\bibfnamefont{Q.~Q.} \bibnamefont{Wang}},
  \bibinfo{author}{\bibfnamefont{A.}~\bibnamefont{Muller}},
  \bibinfo{author}{\bibfnamefont{P.}~\bibnamefont{Bianucci}},
  \bibinfo{author}{\bibfnamefont{E.}~\bibnamefont{Rossi}},
  \bibinfo{author}{\bibfnamefont{Q.~K.} \bibnamefont{Xue}},
  \bibinfo{author}{\bibfnamefont{T.}~\bibnamefont{Takagahara}},
  \bibinfo{author}{\bibfnamefont{C.}~\bibnamefont{Piermarocchi}},
  \bibinfo{author}{\bibfnamefont{A.~H.} \bibnamefont{MacDonald}},
  \bibnamefont{and} \bibinfo{author}{\bibfnamefont{C.~K.} \bibnamefont{Shih}},
  \bibinfo{journal}{Phys. Rev. B} \textbf{\bibinfo{volume}{72}},
  \bibinfo{eid}{035306} (\bibinfo{year}{2005}).

\bibitem[{\citenamefont{Stufler et~al.}(2005)\citenamefont{Stufler, Ester,
  Zrenner, and Bichler}}]{Stufler:121301}
\bibinfo{author}{\bibfnamefont{S.}~\bibnamefont{Stufler}},
  \bibinfo{author}{\bibfnamefont{P.}~\bibnamefont{Ester}},
  \bibinfo{author}{\bibfnamefont{A.}~\bibnamefont{Zrenner}}, \bibnamefont{and}
  \bibinfo{author}{\bibfnamefont{M.}~\bibnamefont{Bichler}},
  \bibinfo{journal}{Phys. Rev. B} \textbf{\bibinfo{volume}{72}},
  \bibinfo{eid}{121301} (\bibinfo{year}{2005}).

\bibitem[{\citenamefont{Vasanelli et~al.}(2002)\citenamefont{Vasanelli,
  Ferreira, and Bastard}}]{Vasanelli:216804}
\bibinfo{author}{\bibfnamefont{A.}~\bibnamefont{Vasanelli}},
  \bibinfo{author}{\bibfnamefont{R.}~\bibnamefont{Ferreira}}, \bibnamefont{and}
  \bibinfo{author}{\bibfnamefont{G.}~\bibnamefont{Bastard}},
  \bibinfo{journal}{Phys. Rev. Lett.} \textbf{\bibinfo{volume}{89}},
  \bibinfo{eid}{216804} (\bibinfo{year}{2002}).

\bibitem[{\citenamefont{Forstner et~al.}(2003)\citenamefont{Forstner, Weber,
  Danckwerts, and Knorr}}]{Forstner:127401}
\bibinfo{author}{\bibfnamefont{J.}~\bibnamefont{Forstner}},
  \bibinfo{author}{\bibfnamefont{C.}~\bibnamefont{Weber}},
  \bibinfo{author}{\bibfnamefont{J.}~\bibnamefont{Danckwerts}},
  \bibnamefont{and} \bibinfo{author}{\bibfnamefont{A.}~\bibnamefont{Knorr}},
  \bibinfo{journal}{Phys. Rev. Lett.} \textbf{\bibinfo{volume}{91}},
  \bibinfo{eid}{127401} (\bibinfo{year}{2003}).

\bibitem[{\citenamefont{Villas-Boas et~al.}(2005)\citenamefont{Villas-Boas,
  Ulloa, and Govorov}}]{Villas-Boas:057404}
\bibinfo{author}{\bibfnamefont{J.~M.} \bibnamefont{Villas-Boas}},
  \bibinfo{author}{\bibfnamefont{S.~E.} \bibnamefont{Ulloa}}, \bibnamefont{and}
  \bibinfo{author}{\bibfnamefont{A.~O.} \bibnamefont{Govorov}},
  \bibinfo{journal}{Phys. Rev. Lett.} \textbf{\bibinfo{volume}{94}},
  \bibinfo{eid}{057404} (\bibinfo{year}{2005}).

\bibitem[{\citenamefont{Bayer et~al.}(2001)\citenamefont{Bayer, Hawrylak,
  Hinzer, Fafard, Korkusinski, Wasilewski, Stern, and Forchel}}]{Bayer:451}
\bibinfo{author}{\bibfnamefont{M.}~\bibnamefont{Bayer}},
  \bibinfo{author}{\bibfnamefont{P.}~\bibnamefont{Hawrylak}},
  \bibinfo{author}{\bibfnamefont{K.}~\bibnamefont{Hinzer}},
  \bibinfo{author}{\bibfnamefont{S.}~\bibnamefont{Fafard}},
  \bibinfo{author}{\bibfnamefont{M.}~\bibnamefont{Korkusinski}},
  \bibinfo{author}{\bibfnamefont{Z.~R.} \bibnamefont{Wasilewski}},
  \bibinfo{author}{\bibfnamefont{O.}~\bibnamefont{Stern}}, \bibnamefont{and}
  \bibinfo{author}{\bibfnamefont{A.}~\bibnamefont{Forchel}},
  \bibinfo{journal}{Science} \textbf{\bibinfo{volume}{291}},
  \bibinfo{pages}{451} (\bibinfo{year}{2001}).

\bibitem[{\citenamefont{Borri et~al.}(2003)\citenamefont{Borri, Langbein,
  Woggon, Schwab, Bayer, Fafard, Wasilewski, and Hawrylak}}]{Borri:267401}
\bibinfo{author}{\bibfnamefont{P.}~\bibnamefont{Borri}},
  \bibinfo{author}{\bibfnamefont{W.}~\bibnamefont{Langbein}},
  \bibinfo{author}{\bibfnamefont{U.}~\bibnamefont{Woggon}},
  \bibinfo{author}{\bibfnamefont{M.}~\bibnamefont{Schwab}},
  \bibinfo{author}{\bibfnamefont{M.}~\bibnamefont{Bayer}},
  \bibinfo{author}{\bibfnamefont{S.}~\bibnamefont{Fafard}},
  \bibinfo{author}{\bibfnamefont{Z.}~\bibnamefont{Wasilewski}},
  \bibnamefont{and} \bibinfo{author}{\bibfnamefont{P.}~\bibnamefont{Hawrylak}},
  \bibinfo{journal}{Phys. Rev. Lett.} \textbf{\bibinfo{volume}{91}},
  \bibinfo{eid}{267401} (\bibinfo{year}{2003}).

\bibitem[{\citenamefont{Talalaev et~al.}(2004)\citenamefont{Talalaev, Tomm,
  Zakharov, Werner, Novikov, and Tonkikh}}]{Talalaev:284}
\bibinfo{author}{\bibfnamefont{V.~G.} \bibnamefont{Talalaev}},
  \bibinfo{author}{\bibfnamefont{J.~W.} \bibnamefont{Tomm}},
  \bibinfo{author}{\bibfnamefont{N.~D.} \bibnamefont{Zakharov}},
  \bibinfo{author}{\bibfnamefont{P.}~\bibnamefont{Werner}},
  \bibinfo{author}{\bibfnamefont{B.~V.} \bibnamefont{Novikov}},
  \bibnamefont{and} \bibinfo{author}{\bibfnamefont{A.~A.}
  \bibnamefont{Tonkikh}}, \bibinfo{journal}{Appl. Phys. Lett.}
  \textbf{\bibinfo{volume}{85}}, \bibinfo{pages}{284} (\bibinfo{year}{2004}).

\bibitem[{\citenamefont{Rontani et~al.}(2004)\citenamefont{Rontani, Amaha,
  Muraki, Manghi, Molinari, Tarucha, and Austing}}]{Rontani:085327}
\bibinfo{author}{\bibfnamefont{M.}~\bibnamefont{Rontani}},
  \bibinfo{author}{\bibfnamefont{S.}~\bibnamefont{Amaha}},
  \bibinfo{author}{\bibfnamefont{K.}~\bibnamefont{Muraki}},
  \bibinfo{author}{\bibfnamefont{F.}~\bibnamefont{Manghi}},
  \bibinfo{author}{\bibfnamefont{E.}~\bibnamefont{Molinari}},
  \bibinfo{author}{\bibfnamefont{S.}~\bibnamefont{Tarucha}}, \bibnamefont{and}
  \bibinfo{author}{\bibfnamefont{D.~G.} \bibnamefont{Austing}},
  \bibinfo{journal}{Phys. Rev. B} \textbf{\bibinfo{volume}{69}},
  \bibinfo{eid}{085327} (\bibinfo{year}{2004}).

\bibitem[{\citenamefont{Krenner
  et~al.}(2005{\natexlab{a}})\citenamefont{Krenner, Sabathil, Clark, Kress,
  Schuh, Bichler, Abstreiter, and Finley}}]{Krenner:057402}
\bibinfo{author}{\bibfnamefont{H.~J.} \bibnamefont{Krenner}},
  \bibinfo{author}{\bibfnamefont{M.}~\bibnamefont{Sabathil}},
  \bibinfo{author}{\bibfnamefont{E.~C.} \bibnamefont{Clark}},
  \bibinfo{author}{\bibfnamefont{A.}~\bibnamefont{Kress}},
  \bibinfo{author}{\bibfnamefont{D.}~\bibnamefont{Schuh}},
  \bibinfo{author}{\bibfnamefont{M.}~\bibnamefont{Bichler}},
  \bibinfo{author}{\bibfnamefont{G.}~\bibnamefont{Abstreiter}},
  \bibnamefont{and} \bibinfo{author}{\bibfnamefont{J.~J.}
  \bibnamefont{Finley}}, \bibinfo{journal}{Phys. Rev. Lett.}
  \textbf{\bibinfo{volume}{94}}, \bibinfo{eid}{057402}
  (\bibinfo{year}{2005}{\natexlab{a}}).

\bibitem[{\citenamefont{Batteh et~al.}(2005)\citenamefont{Batteh, Cheng, Chen,
  Steel, Gammon, Katzer, and Park}}]{Batteh:155327}
\bibinfo{author}{\bibfnamefont{E.~T.} \bibnamefont{Batteh}},
  \bibinfo{author}{\bibfnamefont{J.}~\bibnamefont{Cheng}},
  \bibinfo{author}{\bibfnamefont{G.}~\bibnamefont{Chen}},
  \bibinfo{author}{\bibfnamefont{D.~G.} \bibnamefont{Steel}},
  \bibinfo{author}{\bibfnamefont{D.}~\bibnamefont{Gammon}},
  \bibinfo{author}{\bibfnamefont{D.~S.} \bibnamefont{Katzer}},
  \bibnamefont{and} \bibinfo{author}{\bibfnamefont{D.}~\bibnamefont{Park}},
  \bibinfo{journal}{Phys. Rev. B} \textbf{\bibinfo{volume}{71}},
  \bibinfo{eid}{155327} (\bibinfo{year}{2005}).

\bibitem[{\citenamefont{Bardot et~al.}(2005)\citenamefont{Bardot, Schwab,
  Bayer, Fafard, Wasilewski, and Hawrylak}}]{Bardot:035314}
\bibinfo{author}{\bibfnamefont{C.}~\bibnamefont{Bardot}},
  \bibinfo{author}{\bibfnamefont{M.}~\bibnamefont{Schwab}},
  \bibinfo{author}{\bibfnamefont{M.}~\bibnamefont{Bayer}},
  \bibinfo{author}{\bibfnamefont{S.}~\bibnamefont{Fafard}},
  \bibinfo{author}{\bibfnamefont{Z.}~\bibnamefont{Wasilewski}},
  \bibnamefont{and} \bibinfo{author}{\bibfnamefont{P.}~\bibnamefont{Hawrylak}},
  \bibinfo{journal}{Phys. Rev. B} \textbf{\bibinfo{volume}{72}},
  \bibinfo{eid}{035314} (\bibinfo{year}{2005}).

\bibitem[{\citenamefont{Ortner et~al.}(2005)\citenamefont{Ortner, Yugova, von
  Hogersthal, Larionov, Kurtze, Yakovlev, Bayer, Fafard, Wasilewski, Hawrylak
  et~al.}}]{Ortner:125335}
\bibinfo{author}{\bibfnamefont{G.}~\bibnamefont{Ortner}},
  \bibinfo{author}{\bibfnamefont{I.}~\bibnamefont{Yugova}},
  \bibinfo{author}{\bibfnamefont{G.~B.~H.} \bibnamefont{von Hogersthal}},
  \bibinfo{author}{\bibfnamefont{A.}~\bibnamefont{Larionov}},
  \bibinfo{author}{\bibfnamefont{H.}~\bibnamefont{Kurtze}},
  \bibinfo{author}{\bibfnamefont{D.~R.} \bibnamefont{Yakovlev}},
  \bibinfo{author}{\bibfnamefont{M.}~\bibnamefont{Bayer}},
  \bibinfo{author}{\bibfnamefont{S.}~\bibnamefont{Fafard}},
  \bibinfo{author}{\bibfnamefont{Z.}~\bibnamefont{Wasilewski}},
  \bibinfo{author}{\bibfnamefont{P.}~\bibnamefont{Hawrylak}},
  \bibnamefont{et~al.}, \bibinfo{journal}{Phys. Rev. B}
  \textbf{\bibinfo{volume}{71}}, \bibinfo{eid}{125335} (\bibinfo{year}{2005}).

\bibitem[{\citenamefont{Krenner
  et~al.}(2005{\natexlab{b}})\citenamefont{Krenner, Stufler, Sabathil, Clark,
  Ester, Bichler, Abstreiter, Finley, and Zrenner}}]{Krenner:184}
\bibinfo{author}{\bibfnamefont{H.~J.} \bibnamefont{Krenner}},
  \bibinfo{author}{\bibfnamefont{S.}~\bibnamefont{Stufler}},
  \bibinfo{author}{\bibfnamefont{M.}~\bibnamefont{Sabathil}},
  \bibinfo{author}{\bibfnamefont{E.~C.} \bibnamefont{Clark}},
  \bibinfo{author}{\bibfnamefont{P.}~\bibnamefont{Ester}},
  \bibinfo{author}{\bibfnamefont{M.}~\bibnamefont{Bichler}},
  \bibinfo{author}{\bibfnamefont{G.}~\bibnamefont{Abstreiter}},
  \bibinfo{author}{\bibfnamefont{J.~J.} \bibnamefont{Finley}},
  \bibnamefont{and} \bibinfo{author}{\bibfnamefont{A.}~\bibnamefont{Zrenner}},
  \bibinfo{journal}{New Journal of Physics} \textbf{\bibinfo{volume}{7}},
  \bibinfo{pages}{184} (\bibinfo{year}{2005}{\natexlab{b}}).

\bibitem[{\citenamefont{Unold et~al.}(2005)\citenamefont{Unold, Mueller,
  Lienau, Elsaesser, and Wieck}}]{Unold:137404}
\bibinfo{author}{\bibfnamefont{T.}~\bibnamefont{Unold}},
  \bibinfo{author}{\bibfnamefont{K.}~\bibnamefont{Mueller}},
  \bibinfo{author}{\bibfnamefont{C.}~\bibnamefont{Lienau}},
  \bibinfo{author}{\bibfnamefont{T.}~\bibnamefont{Elsaesser}},
  \bibnamefont{and} \bibinfo{author}{\bibfnamefont{A.~D.} \bibnamefont{Wieck}},
  \bibinfo{journal}{Phys. Rev. Lett.} \textbf{\bibinfo{volume}{94}},
  \bibinfo{eid}{137404} (\bibinfo{year}{2005}).

\bibitem[{\citenamefont{Stinaff et~al.}(2006)\citenamefont{Stinaff, Scheibner,
  Bracker, Ponomarev, Korenev, Ware, Doty, Reinecke, and Gammon}}]{Stinaff:636}
\bibinfo{author}{\bibfnamefont{E.~A.} \bibnamefont{Stinaff}},
  \bibinfo{author}{\bibfnamefont{M.}~\bibnamefont{Scheibner}},
  \bibinfo{author}{\bibfnamefont{A.~S.} \bibnamefont{Bracker}},
  \bibinfo{author}{\bibfnamefont{I.~V.} \bibnamefont{Ponomarev}},
  \bibinfo{author}{\bibfnamefont{V.~L.} \bibnamefont{Korenev}},
  \bibinfo{author}{\bibfnamefont{M.~E.} \bibnamefont{Ware}},
  \bibinfo{author}{\bibfnamefont{M.~F.} \bibnamefont{Doty}},
  \bibinfo{author}{\bibfnamefont{T.~L.} \bibnamefont{Reinecke}},
  \bibnamefont{and} \bibinfo{author}{\bibfnamefont{D.}~\bibnamefont{Gammon}},
  \bibinfo{journal}{Science} \textbf{\bibinfo{volume}{311}},
  \bibinfo{pages}{636} (\bibinfo{year}{2006}).

\bibitem[{\citenamefont{Robledo et~al.}(2008)\citenamefont{Robledo, Elzerman,
  Jundt, Atature, Hogele, Falt, and Imamoglu}}]{Robledo2008}
\bibinfo{author}{\bibfnamefont{L.}~\bibnamefont{Robledo}},
  \bibinfo{author}{\bibfnamefont{J.}~\bibnamefont{Elzerman}},
  \bibinfo{author}{\bibfnamefont{G.}~\bibnamefont{Jundt}},
  \bibinfo{author}{\bibfnamefont{M.}~\bibnamefont{Atature}},
  \bibinfo{author}{\bibfnamefont{A.}~\bibnamefont{Hogele}},
  \bibinfo{author}{\bibfnamefont{S.}~\bibnamefont{Falt}}, \bibnamefont{and}
  \bibinfo{author}{\bibfnamefont{A.}~\bibnamefont{Imamoglu}},
  \bibinfo{journal}{Science} \textbf{\bibinfo{volume}{320}},
  \bibinfo{pages}{772} (\bibinfo{year}{2008}).

\bibitem[{\citenamefont{Bracker et~al.}(2006)\citenamefont{Bracker, Scheibner,
  Doty, Stinaff, Ponomarev, Kim, Whitman, Reinecke, and Gammon}}]{Bracker06}
\bibinfo{author}{\bibfnamefont{A.~S.} \bibnamefont{Bracker}},
  \bibinfo{author}{\bibfnamefont{M.}~\bibnamefont{Scheibner}},
  \bibinfo{author}{\bibfnamefont{M.~F.} \bibnamefont{Doty}},
  \bibinfo{author}{\bibfnamefont{E.~A.} \bibnamefont{Stinaff}},
  \bibinfo{author}{\bibfnamefont{I.~V.} \bibnamefont{Ponomarev}},
  \bibinfo{author}{\bibfnamefont{J.~C.} \bibnamefont{Kim}},
  \bibinfo{author}{\bibfnamefont{L.~J.} \bibnamefont{Whitman}},
  \bibinfo{author}{\bibfnamefont{T.~L.} \bibnamefont{Reinecke}},
  \bibnamefont{and} \bibinfo{author}{\bibfnamefont{D.}~\bibnamefont{Gammon}},
  \bibinfo{journal}{Appl. Phys. Lett.} \textbf{\bibinfo{volume}{89}},
  \bibinfo{pages}{233110} (\bibinfo{year}{2006}).

\bibitem[{\citenamefont{Villas-Bôas et~al.}(2004)\citenamefont{Villas-Bôas,
  Gororov, and Ulloa}}]{Villas04}
\bibinfo{author}{\bibfnamefont{J.}~\bibnamefont{Villas-Bôas}},
  \bibinfo{author}{\bibfnamefont{A.}~\bibnamefont{Gororov}}, \bibnamefont{and}
  \bibinfo{author}{\bibfnamefont{S.}~\bibnamefont{Ulloa}},
  \bibinfo{journal}{Phys. Rev. B} \textbf{\bibinfo{volume}{69}},
  \bibinfo{pages}{125342} (\bibinfo{year}{2004}).

\bibitem[{\citenamefont{Calarco et~al.}(2003)\citenamefont{Calarco, Datta,
  Fedichev, Pazy, and Zoller}}]{Calarco03}
\bibinfo{author}{\bibfnamefont{T.}~\bibnamefont{Calarco}},
  \bibinfo{author}{\bibfnamefont{A.}~\bibnamefont{Datta}},
  \bibinfo{author}{\bibfnamefont{P.}~\bibnamefont{Fedichev}},
  \bibinfo{author}{\bibfnamefont{E.}~\bibnamefont{Pazy}}, \bibnamefont{and}
  \bibinfo{author}{\bibfnamefont{P.}~\bibnamefont{Zoller}},
  \bibinfo{journal}{Phys. Rev. A} \textbf{\bibinfo{volume}{68}},
  \bibinfo{pages}{012310} (\bibinfo{year}{2003}).

\bibitem[{\citenamefont{Sakurai}(1994)}]{Sakurai94}
\bibinfo{author}{\bibfnamefont{J.~J.} \bibnamefont{Sakurai}},
  \emph{\bibinfo{title}{Modern Quantum Mechanics}} (\bibinfo{publisher}{Addison
  Wesley}, \bibinfo{year}{1994}).

\bibitem[{\citenamefont{Rau and Zhao}(2003)}]{Zhao03}
\bibinfo{author}{\bibfnamefont{A.~R.~P.} \bibnamefont{Rau}} \bibnamefont{and}
  \bibinfo{author}{\bibfnamefont{W.}~\bibnamefont{Zhao}},
  \bibinfo{journal}{Phys. Rev. A} \textbf{\bibinfo{volume}{68}},
  \bibinfo{pages}{052102} (\bibinfo{year}{2003}).

\bibitem[{\citenamefont{Villas-Bôas et~al.}(2007)\citenamefont{Villas-Bôas,
  Ulloa, and Gororov}}]{Villas07}
\bibinfo{author}{\bibfnamefont{J.}~\bibnamefont{Villas-Bôas}},
  \bibinfo{author}{\bibfnamefont{S.}~\bibnamefont{Ulloa}}, \bibnamefont{and}
  \bibinfo{author}{\bibfnamefont{A.}~\bibnamefont{Gororov}},
  \bibinfo{journal}{Phys. Rev. B} \textbf{\bibinfo{volume}{75}},
  \bibinfo{pages}{155334} (\bibinfo{year}{2007}).

\bibitem[{\citenamefont{Bonadeo et~al.}(1998)\citenamefont{Bonadeo, Erland,
  Gammon, Park, Katzer, and Steel}}]{Gammon98}
\bibinfo{author}{\bibfnamefont{N.~H.} \bibnamefont{Bonadeo}},
  \bibinfo{author}{\bibfnamefont{J.}~\bibnamefont{Erland}},
  \bibinfo{author}{\bibfnamefont{D.}~\bibnamefont{Gammon}},
  \bibinfo{author}{\bibfnamefont{D.}~\bibnamefont{Park}},
  \bibinfo{author}{\bibfnamefont{D.~S.} \bibnamefont{Katzer}},
  \bibnamefont{and} \bibinfo{author}{\bibfnamefont{D.~G.} \bibnamefont{Steel}},
  \bibinfo{journal}{Science} \textbf{\bibinfo{volume}{282}},
  \bibinfo{pages}{1473} (\bibinfo{year}{1998}).

\bibitem[{\citenamefont{Chen et~al.}(2001)\citenamefont{Chen, Piermarocchi, and
  Sham}}]{Chen01}
\bibinfo{author}{\bibfnamefont{P.}~\bibnamefont{Chen}},
  \bibinfo{author}{\bibfnamefont{C.}~\bibnamefont{Piermarocchi}},
  \bibnamefont{and} \bibinfo{author}{\bibfnamefont{L.~J.} \bibnamefont{Sham}},
  \bibinfo{journal}{Phys. Rev. Lett.} \textbf{\bibinfo{volume}{87}},
  \bibinfo{pages}{67401} (\bibinfo{year}{2001}).

\bibitem[{\citenamefont{Takagahara}(2002)}]{Takagahara02}
\bibinfo{author}{\bibfnamefont{T.}~\bibnamefont{Takagahara}},
  \bibinfo{journal}{Phys. Stat. Sol.} \textbf{\bibinfo{volume}{234}},
  \bibinfo{pages}{115} (\bibinfo{year}{2002}).

\bibitem[{\citenamefont{Negoita et~al.}(1999)\citenamefont{Negoita, Snoke, and
  Eberl}}]{Negoita99}
\bibinfo{author}{\bibfnamefont{V.}~\bibnamefont{Negoita}},
  \bibinfo{author}{\bibfnamefont{D.~W.} \bibnamefont{Snoke}}, \bibnamefont{and}
  \bibinfo{author}{\bibfnamefont{K.}~\bibnamefont{Eberl}},
  \bibinfo{journal}{Phys. Rev. B} \textbf{\bibinfo{volume}{60}},
  \bibinfo{pages}{2661} (\bibinfo{year}{1999}).

\bibitem[{\citenamefont{Tackeuchi et~al.}(2000)\citenamefont{Tackeuchi, Kuroda,
  Mase, Nakata, and Yokoyama}}]{Tackeuchi00}
\bibinfo{author}{\bibfnamefont{A.}~\bibnamefont{Tackeuchi}},
  \bibinfo{author}{\bibfnamefont{T.}~\bibnamefont{Kuroda}},
  \bibinfo{author}{\bibfnamefont{K.}~\bibnamefont{Mase}},
  \bibinfo{author}{\bibfnamefont{Y.}~\bibnamefont{Nakata}}, \bibnamefont{and}
  \bibinfo{author}{\bibfnamefont{N.}~\bibnamefont{Yokoyama}},
  \bibinfo{journal}{Phys. Rev. B} \textbf{\bibinfo{volume}{62}},
  \bibinfo{pages}{1568} (\bibinfo{year}{2000}).

\bibitem[{\citenamefont{Emary and Sham}(2007)}]{Emary07}
\bibinfo{author}{\bibfnamefont{C.}~\bibnamefont{Emary}} \bibnamefont{and}
  \bibinfo{author}{\bibfnamefont{L.}~\bibnamefont{Sham}},
  \bibinfo{journal}{Phys. Rev. B} \textbf{\bibinfo{volume}{75}},
  \bibinfo{pages}{125317} (\bibinfo{year}{2007}).

\end{thebibliography}

\end{document}